\documentclass[letters,fleqn,usenatbib,useAMS]{mnras}

\usepackage[T1]{fontenc}
\usepackage{ae,aecompl}

\usepackage{graphicx}	
\usepackage{amsmath}	
\usepackage{amssymb}	

\newcommand{\bz}{\left< B_{\rm z} \right>}

\title[Pulsating stars]{Searching for magnetic fields in pulsating A-type stars:
the discovery of a strong field in the probable
$\delta$\,Sct star HD\,340577 and a null result for the $\gamma$\,Dor star HR\,8799}

\stepcounter{footnote}

\author[Hubrig et al.]{
S.~Hubrig$^{1}$,
S.~P.~J\"arvinen$^{1}$,
J.~D.~Alvarado-G\'omez$^{1}$,
I.~Ilyin$^{1}$,
M.~Sch\"oller$^{2}$
\\
$^{1}${Leibniz-Institut f\"ur Astrophysik Potsdam (AIP), An der Sternwarte~16, 14482~Potsdam, Germany} \\
$^{2}${European Southern Observatory, Karl-Schwarzschild-Str.~2, 85748 Garching, Germany}
}

\date{Accepted XXX. Received YYY; in original form ZZZ}

\pubyear{2023}

\begin{document}
\label{firstpage}
\pagerange{\pageref{firstpage}--\pageref{lastpage}}
\maketitle

\begin{abstract}
Numerous $\delta$\,Sct and $\gamma$\,Dor pulsators are identified in the 
region of the Hertzsprung-Russell diagram that is occupied by chemically 
peculiar magnetic Ap stars. The connection between $\delta$\,Sct and 
$\gamma$\,Dor pulsations and the magnetic field in Ap stars is however not 
clear: theory suggests for magnetic Ap stars some critical field strengths 
for pulsation  mode suppression by computing 
the magnetic damping effect for selected $p$ and $g$ modes. 
To test these theoretical considerations, we obtained PEPSI spectropolarimetric
snapshots of the typical Ap star HD\,340577,
for which $\delta$\,Sct-like pulsations were recently detected in TESS data,
and the  $\gamma$\,Dor pulsator HR\,8799, which is a remarkable system with multiple planets
and two debris disks.
Our measurements reveal the presence of a magnetic field with a strength of several hundred Gauss in
HD\,340577. The measured mean longitudinal field would be the strongest 
field measured so far in a $\delta$\,Sct star if the pulsational character of HD\,340577 is confirmed
spectroscopically.
No magnetic field is detected in HR\,8799.
\end{abstract}




\section{Introduction}
\label{sec:intro}

About 15\% of the upper main-sequence stars between spectral types early B 
and early F are magnetic chemically peculiar stars, characterized by 
atmospheric chemical abundances that differ significantly from the solar 
pattern. The observed peculiarity is believed to be produced by selective 
processes, i.e.\ radiative levitation and gravitational settling, in the 
presence of a magnetic field. These chemically peculiar magnetic stars, 
generally called Ap and Bp stars, are well-known for displaying a non-uniform 
distribution of chemical elements in the form of spots and patches and usually
exhibit strong globally organized magnetic fields 
that range from tens of Gauss to several tens of kG 
(e.g.\ \citealt{HubrigSchoeller2021}). 

In the region of the A- and F-type stars, where most of the magnetic Ap stars 
are situated, also numerous $\delta$\,Sct and $\gamma$\,Dor pulsators were 
identified using photometric surveys. The $\delta$\,Sct stars are located on  
and near the main sequence within the classical instability strip in the 
Hertzsprung-Russell diagram. They have pulsation periods between about tens 
of minutes and 8\,h, corresponding to radial and nonradial pressure ($p$) and mixed 
modes of low radial order. The $\gamma$\,Dor variables are late-A and early-F
type stars showing gravity ($g$) mode pulsations with periods between 0.3 and 3 
days \citep{Balona1994}. 
They are located between the $\delta$\,Sct stars and solar-like oscillators, 
with numerous hybrid $\delta$\,Sct and $\gamma$\,Dor stars pulsating in 
$p$ and $g$ modes 
\citep{Grigahcene2005}. 
The connection between $\delta$\,Sct and $\gamma$\,Dor pulsations and 
magnetic fields in Ap stars is however not clear. Previous theoretical work 
(e.g.\ \citealt{Saio2005}) 
suggested that magnetic Ap stars should not show $\delta$\,Sct or 
$\gamma$\,Dor pulsations. Most recently, 
\citet{Murphy2020}
reported on some critical field strengths for mode suppression by computing 
the magnetic damping effect for selected $p$ and $g$ modes. The authors 
found that magnetic field strengths less than 1\,kG do not suppress 
low-overtone $p$ modes. The high-order $g$ modes typical for $\gamma$\,Dor 
stars are expected to be damped by dipole fields stronger than 1–4\,kG, while 
the low-order $g$ modes with frequencies similar to the fundamental mode are 
not suppressed.

The first discovery of chemical peculiarities in $\delta$\,Sct stars, similar 
to those in magnetic Ap stars, was announced by 
\citet{Koen2001},
who found $\delta$\,Sct pulsations in HD\,21190, classified as 
F2III\,SrEu(Si). The presence of a magnetic field in this star was reported 
by \citet{Kurtz2008}
and later confirmed by 
\citet{Jarvinen2018}. 
Very few other discoveries of magnetic fields in $\delta$\,Sct stars
were reported in the past, usually rather weak magnetic fields,
down to below 1\,Gauss (e.g.,
\citealt{Neiner2017};
\citealt{Zwintz2020}). 
None of these previously studied $\delta$\,Sct pulsators with weak fields 
presented chemical peculiarities typical for Ap stars.
As an exceptional case, \citet{Murphy2020} reported on the detection of the hybrid 
pulsator KIC 11296437, showing 
both low-overtone and high-overtone $p$ modes with a mean magnetic 
field modulus of $2.8\pm0.5$\,kG. This strong magnetic field was estimated using the empirical 
relation by \citet{MathysLanz1992} based on the ratio of the equivalent widths of the 
lines \ion{Fe}{ii}~6147.7\,\AA{} and \ion{Fe}{ii}~6149.2\,\AA{}.
No spectropolarimetric measurements were carried out yet for this hybrid pulsator.
The scarcity of previous detections of magnetic fields in Ap stars with $\delta$\,Sct pulsations 
indicates that such targets are extremely rare.

Quite unexpectedly, \citet{Mathys2022} reported on  the detection of nine definite and five
candidate $\delta$\,Sct stars in their systematic search for periodicities in typical 
chemically peculiar Ap stars based on an analysis of Transiting Exoplanet 
Survey Satellite (TESS) data. 
Importantly, no Ap star with definite $\gamma$\,Dor pulsations was detected in this survey. 
On the other hand, according to \citet{Li2020}, about 10\% of photometrically 
studied $\gamma$\,Dor targets show rotational modulation expected to be caused by a surface inhomogeneous chemical 
element distribution.
As this fraction roughly corresponds to the occurrence of magnetic Ap stars among A-type stars, it is expected that 
$\gamma$\,Dor stars also possess magnetic fields.
Unfortunately, no systematic search for magnetic fields was carried out in $\gamma$\,Dor stars in the past, so that 
their magnetic nature appears currently elusive.

To test previous theoretical and observational considerations, 
we used the Potsdam Echelle Polarimetric and 
Spectroscopic Instrument (PEPSI; \citealt{Strassmeier2015}) installed at 
the $2\times8.4$\,m Large Binocular Telescope (LBT) to observe HD\,340577
with classification A3p SrCrEu \citep{Renson2009}, reported by \citet{Mathys2022} to exhibit
$\delta$\,Sct-like pulsations,  and the A5 $\gamma$\,Dor star 
HR\,8799 (e.g.\ \citealt{Sodor2014}). 
The $\gamma$\,Dor star HR\,8799 with a mass of about 1.5\,$M_{\odot}$ is one of the rare systems 
in which multiple planets have been directly imaged (e.g.\ \citealt{Marois2008}).
Its architecture is strikingly similar to that 
of the solar system, with the four giant planets surrounding a warm dust belt analogous to the Asteroid Belt, 
and themselves being surrounded by a cold dust belt analog to the Kuiper Belt (e.g.\ 
\citealt{Faramaz2021}).  HR\,8799 is therefore of special 
interest to search for the presence of a magnetic field to improve our understanding of the magnetospheric 
interactions between the host star and the planets.
HR\,8799 is also identified as a $\lambda$\,Bootis star. One of the scenarios to explain 
$\lambda$\,Bootis-like chemical abundances suggests that planetary  bodies  could  perturb  the  orbits  of
comets and volatile-rich objects, likely sending them toward the star and possibly polluting it
\citep{Gray2002}.

In the following, we describe  the PEPSI spectropolarimetric observations, the methodology of our 
analysis and present the results of the magnetic field measurements (Sect.~2).
In Sect.~3  we discuss the implication of our results on future studies of A-type stars with different 
pulsational characteristics.

\section{Observations and magnetic field measurements}
\label{sec:specpol}

Both stars were observed with PEPSI on the LBT on 2022 September~15. 
A description of the spectropolarimetric mode of PEPSI and the data reduction
are discussed in the works by \citet{Strassmeier2023} and \citet{Hubrig2021}.
Three grisms per PEPSI spectrograph arm cover the wavelength range from 3837 to 9067\,\AA{}.
All polarimetric spectra have a fixed spectral resolution of $R\sim130\,000$ corresponding to 0.06\,\AA{}
at 7600\,\AA{}. We used cross
dispersers CD\,II covering 4236--4770\,\AA{} and CD\,IV covering 5391--6289\,\AA{}.
With an exposure time of about 5\,min for one retarder angle we could achieve for HD\,340577 a signal-to-noise 
ratio of about 250 whereas for the bright star HR\,8799 we needed  
only 2.3\,min for one retarder angle to achieve a signal-to-noise ratio of 630.

Similar to our previous studies, to measure the mean longitudinal magnetic field and to increase
the signal-to-noise ratio, we employed the least-squares deconvolution (LSD) technique.
The details of this technique, 
as well as how the LSD Stokes~$I$, Stokes~$V$, and diagnostic null spectra are calculated, were 
presented by \citet{Donati1997}.
To evaluate, whether the detected features are spurious or definite detections, 
we followed the generally adopted procedure to use the false alarm probability (${\rm FAP}$) based on 
reduced $\chi^{2}$ test statistics \citep{Donati1992}:
the presence of a Zeeman signature
is considered as a definite detection 
if ${\rm FAP} \leq 10^{-5}$,
as a marginal detection if  $10^{-5}<{\rm FAP}\leq 10^{-3}$,
and as a non-detection if ${\rm FAP}>10^{-3}$.

\begin{figure}
\centering 
\includegraphics[width=0.840\columnwidth]{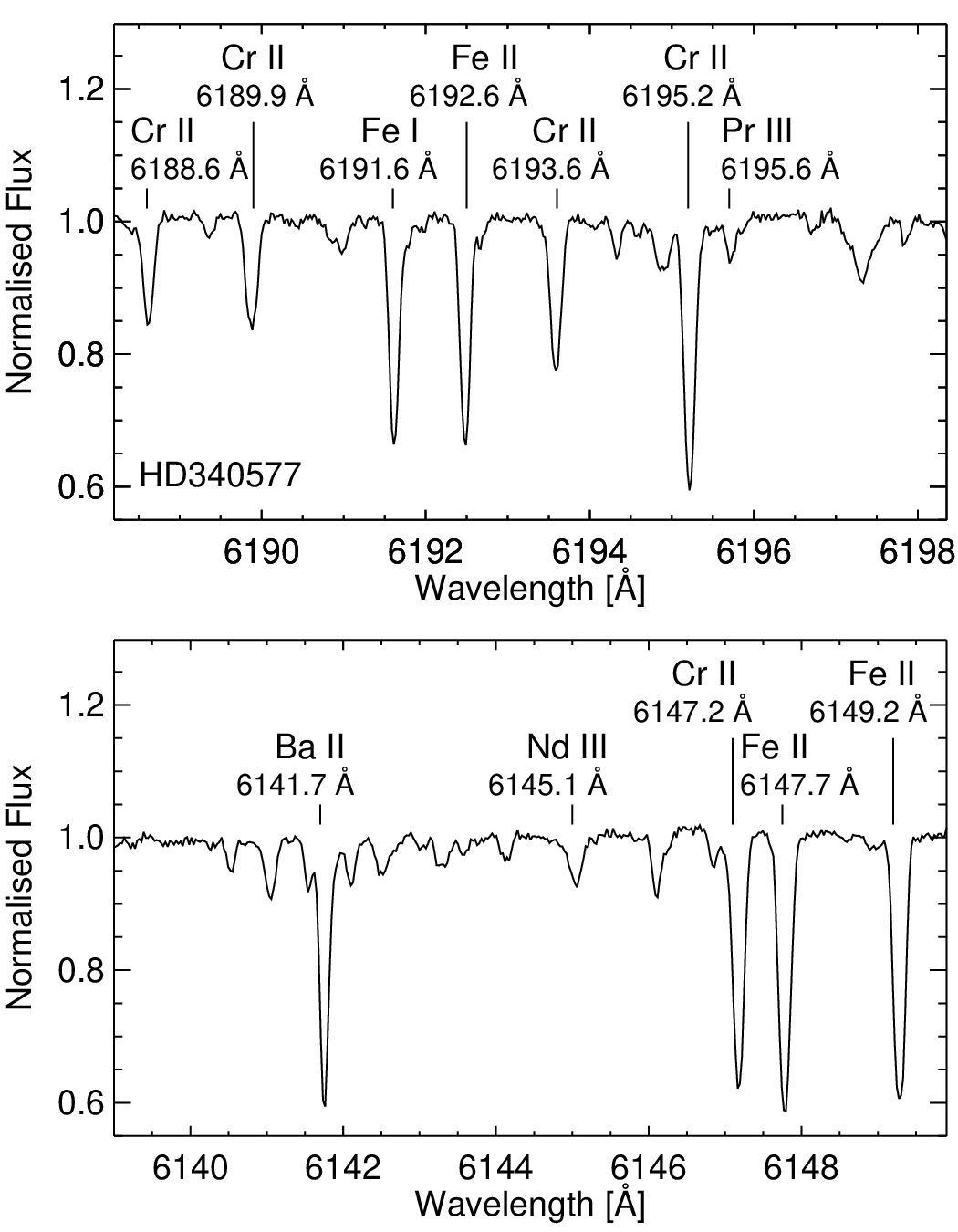} 
\caption{
Portions of the PEPSI spectrum of HD\,340577 around the expected positions for \ion{Nd}{iii}~6145.1\,\AA{}
(bottom panel) and \ion{Pr}{iii}~6195.6\,\AA{} (top panel). A few other spectral lines that 
are typical of A-type stars are also identified. 
}
\label{fig:ree-hd}
\end{figure}

\begin{figure}
\centering 
\includegraphics[width=0.840\columnwidth]{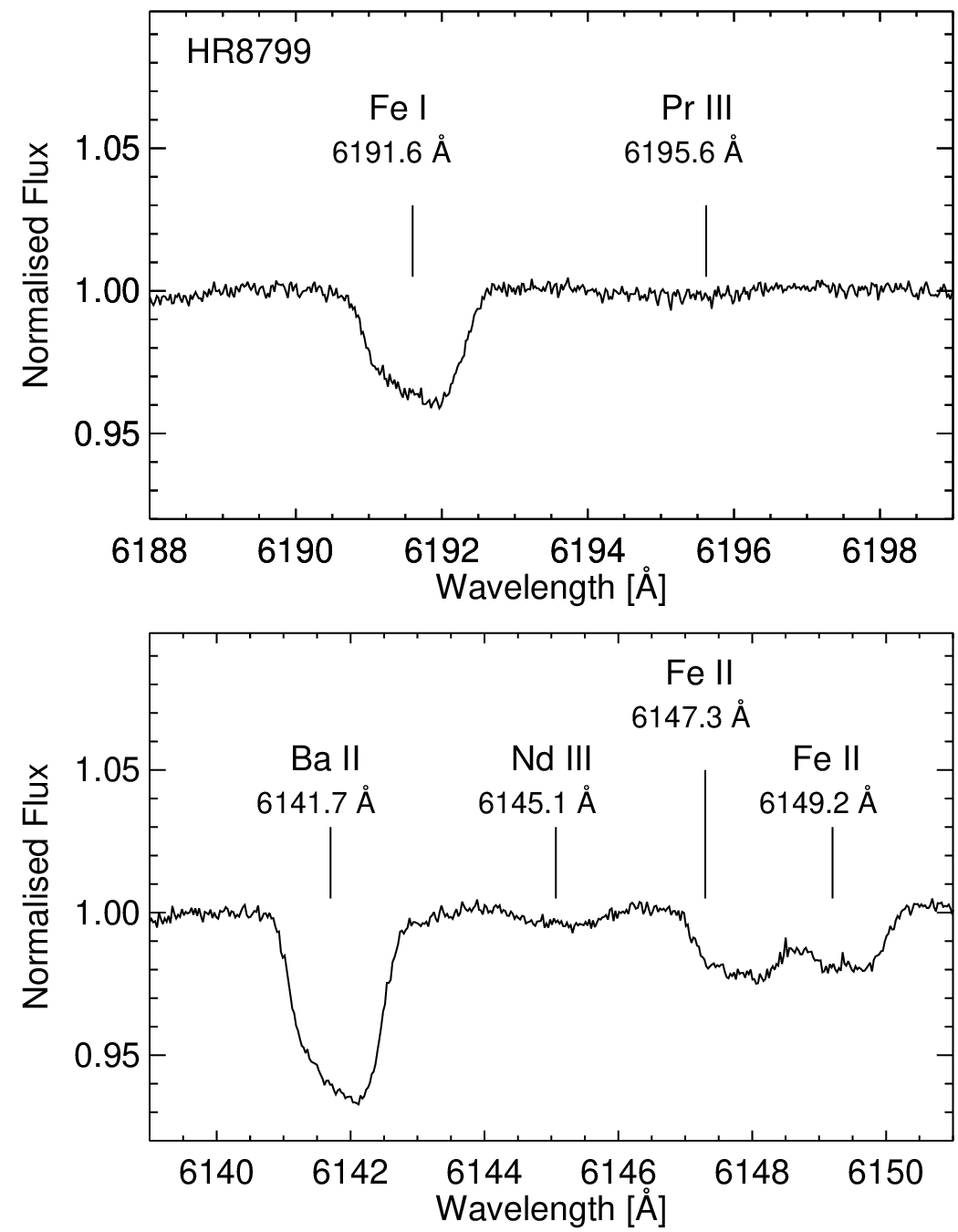} 
\caption{
As Fig.~\ref{fig:ree-hd} for the $\gamma$\,Dor star HR\,8799.
}
\label{fig:ree-hr}
\end{figure}

The inspection of the PEPSI Stokes~$I$ spectra of both, the Ap star HD\,340577 and the 
$\gamma$\,Dor star HR\,8799 reveals that the spectral lines belonging to the rare earth 
elements (REE) Pr and Nd are very weak, in 
contrast to those usually observed in magnetic Ap stars (see for comparison e.g.\ 
fig.~1 in \citealt{Hubrig2018}).
In the $\gamma$\,Dor star HR\,8799, 
even one of the strongest Nd lines, the \ion{Nd}{iii}~6145.1\,\AA{} line, can hardly be identified 
in the spectrum. In Figs.~\ref{fig:ree-hd} and \ref{fig:ree-hr} we present portions of the PEPSI 
Stokes~$I$ spectra of both stars around the expected positions for \ion{Nd}{iii}~6145.1\,\AA{} 
and \ion{Pr}{iii}~6195.6\,\AA{}.

Magnetic Ap stars also generally display non-uniform horizontal 
and vertical distributions of chemical elements with certain elements showing a preferential surface 
concentration close to the magnetic poles while other elements concentrate closer to the magnetic equator regions.
Combining lines belonging to different elements
altogether, as is frequently done with the LSD technique, may lead
to the dilution of the magnetic signal or even to its (partial) cancellation,
if enhancements of different elements occur in regions of opposite
magnetic polarity.  Therefore,  several line lists corresponding to different elements in different 
ionization states are usually created to investigate the potential impact of the 
non-uniform element distribution.
Line masks corresponding to the spectral types of both stars 
were constructed using the Vienna Atomic Line Database \citep[VALD3;][]{Kupka2011}.
We created for HD\,340577 in total seven masks:
one line mask each for the neutral \ion{Fe}{i} and \ion{Mn}{i}
and for the five first ions
\ion{Ti}{ii}, \ion{Cr}{ii}, \ion{Fe}{ii}, \ion{Ba}{ii}, and the REE \ion{Eu}{ii}.
All of them contain the least blended best identified spectral lines.
For HR\,8799, we created five line masks,
for the neutral \ion{Ca}{i} and \ion{Fe}{i},
the first ions \ion{Ti}{ii} and \ion{Fe}{ii},
and for \ion{Cr}{i} and \ion{Cr}{ii} together.
We had to combine \ion{Cr}{i} and \ion{Cr}{ii}
due to the small number of identified unblended \ion{Cr}{i/ii} lines.

\begin{figure*}
\centering 
\includegraphics[width=0.840\textwidth]{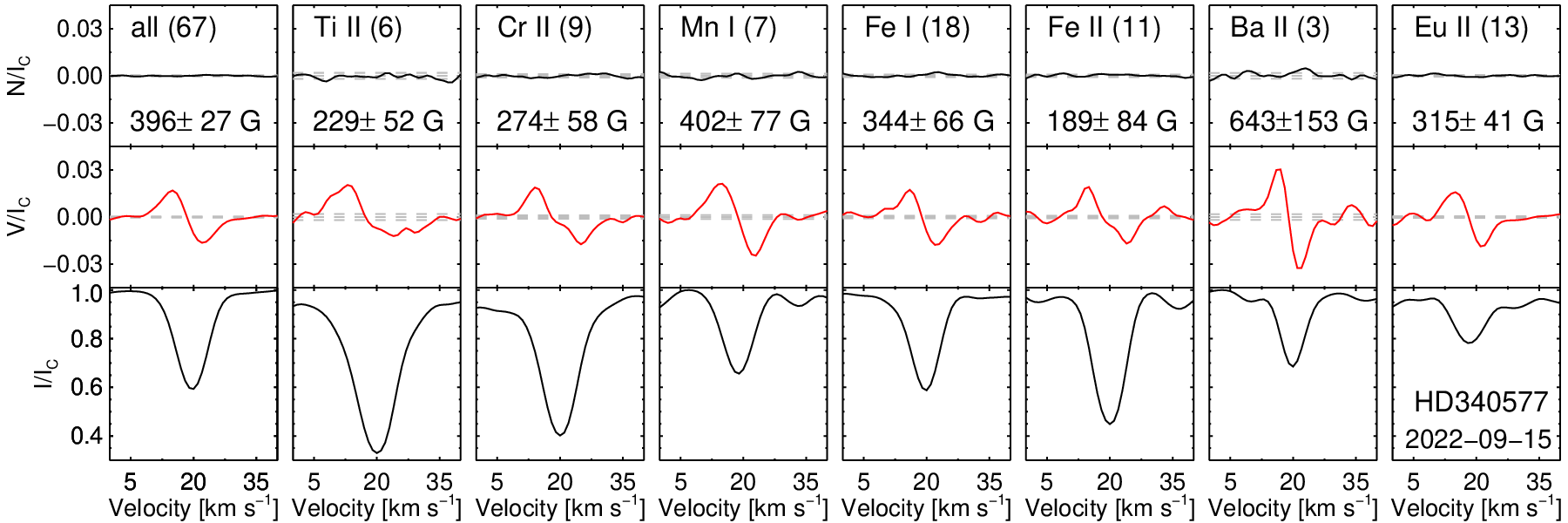} 
\caption{
LSD Stokes~$I$, $V$, and diagnostic null $N$ spectra (from bottom to top)
calculated for HD\,340577,
for the mask with all lines and the masks for individual ions.
The numbers in brackets relate to the number of lines used in each individual mask.
The grey horizontal lines in the Stokes~$V$ and $N$ spectra
indicate the $\pm1\sigma$ error bars.
Stokes~$V$ spectra are highlighted in red.
}
\label{fig:scuti}
\end{figure*}

\begin{figure*}
\centering 
\includegraphics[width=0.845\textwidth]{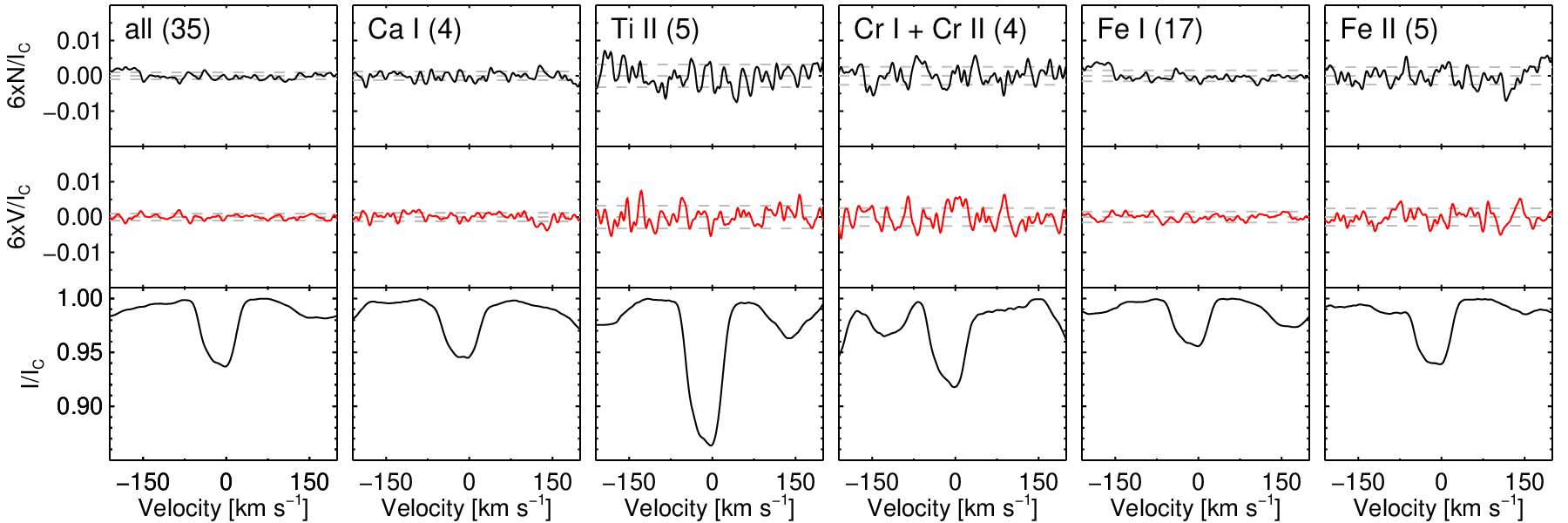} 
\caption{
As Fig.~\ref{fig:scuti} for the $\gamma$\,Dor star HR\,8799.
The  Stokes~$V$ and $N$ spectra are magnified for better visibility.
}
\label{fig:dor}
\end{figure*}

The  results  of  the  LSD  magnetic field  measurements using
different line masks are presented for the Ap star HD\,340577 in Fig.~\ref{fig:scuti} and 
for the $\gamma$\,Dor star HR\,8799 in Fig.~\ref{fig:dor}. 
In these figures, the first plots on the left show our measurements using a mask containing the lines of all elements.
In total, 67 spectral lines were used in the line list for
the measurements of HD\,340577 whereas the number of lines used for the faster rotating HR\,8799 was only 35.
Our measurements show that HD\,340577 possesses a rather strong mean longitudinal magnetic
field, up to more than 600\,G measured in the \ion{Ba}{ii} lines. The fact that the measurements using individual
masks constructed for different ions show
differences in the field strengths, suggests that the lines belonging to the elements Ti and Cr with 
the lowest field strengths form at some distance from the magnetic pole whereas other elements form closer 
to the magnetic pole. The difference in the field strengths between the measurements using 
\ion{Fe}{i} and \ion{Fe}{ii} masks might indicate a vertical Fe stratification.
For all measurements the FAP for the field detection in HD\,340577 is less than
$<10^{-8}$.

No magnetic field is detected in our measurements using any of the six line masks constructed 
for the $\gamma$\,Dor star HR\,8799.
Our simulations of a Zeeman feature with an amplitude corresponding to the noise level in the measurements 
using the line mask containing all lines indicates a detection limit 
of just 9\,Gauss. For all elements the line profiles centered at $V_{\rm R} = -12.4$\,km\,s$^{-1}$ in the 
Stokes~$I$ spectra show asymmetric profiles,
probably caused by the presence of  $\gamma$\,Dor-like pulsations. 
The early GAIA DR3 \citep{Gaia2016,Gaia2021}
places the closest stellar objects on 
the sky at 57.8\,arcsec (TYC 1718-851-1,
G0, $G = 9.46$\,mag, $V_{\rm R} = 17.45$\,km\,s$^{-1}$) 
and 188.4\,arcsec (TIC 315386410, Horizontal Branch Star, 
$G = 15.73$\,mag, $V_{\rm R} = -145.2$\,km\,s$^{-1}$).

Infrared observations of HR\,8799 indicate the presence of an inner 
warm disk in the system \citep{Reidemeister2009,Su2009,Su2013}. 
This disk is unresolved and its properties are still poorly constrained.
They have been 
inferred based only on associated SED fits and dynamical stability arguments. These models place 
the inner edge of the warm disk somewhere in the region between $2-6$\,au 
(see \citealt{KennedyWyatt2014}; \citealt{Contro2016}). 

The Stokes~$I$ spectra of HR\,8799 presented in Fig.~\ref{fig:dor} show that,
apart from the LSD Stokes~$I$ absorption profiles associated with the star,
other absorption features, depending on the line mask used in the LSD retrieval, are 
detectable on the blue and red sides around $130-150$\,km\,s$^{-1}$ in absolute values.
It is not entirely clear what the origin of these secondary signatures is.
Their appearance can be expected in the LSD technique if the lines selected for the line 
masks have close neighbouring
spectral lines or due to the small number of
lines in the line masks.
Importantly,  HR\,8799 was detected at soft X-ray energies with the Chandra ACIS-S detector in a 
10\,ks exposure by \citet{RobradeSchmitt2010}.
Since the $\lambda$\,Boo chemical abundances are usually explained involving an accretion scenario and 
stellar accretion signatures generally yield rather soft X-ray spectra,
our considerations of the presence of an accreting warm gas disk around HR\,8799
may probably be discussed as a working scenario.

\section{Discussion}
\label{sec:disc}

So far, no systematic search for magnetic fields was carried out in $\delta$\,Sct and
$\gamma$\,Dor stars. Using a snapshot of the Ap star 
HD\,340577,
for which $\delta$\,Sct-like pulsations were recently detected in TESS data,
with the PEPSI spectropolarimeter
we discover a quite strong mean longitudinal magnetic field
with $\bz=396\pm27$\,G using a line mask constructed for various elements. As of today, HD\,340577 possesses
the strongest magnetic field among the previously reported magnetic field detections in stars with
$\delta$\,Sct-like pulsations.
While no signs for multiplicity have been detected so far for HD\,340577,
it is still possible that the recorded  $\delta$\,Sct-like pulsations originate in
an undetected companion.
Obviously, future spectroscopic time series are
needed to decide if the pulsations indeed originate in the Ap star.
Assuming a dipole field configuration we estimate for HD\,340577 a minimum dipole strength 
of $\sim1.2$\,kG using the relation
$B_{\rm d} \ge 3 \left| \left<B_{\rm z}\right>_{\rm max} \right|$ \citep{Babcock1958}. Our results are just
slightly outside of the reported boundary of 1\,kG,
for which previous model calculations indicate that low-overtone $p$ modes are not suppressed.

As is mentioned in Sect.~\ref{sec:intro},
\citet{Murphy2020} detected the hybrid pulsator 
KIC~11296437 with a considerably stronger magnetic field, but this detection is still pending a confirmation
by spectropolarimetric observations. 
The next strongest field, $\bz=350\pm100$\,G was reported for 
the $\delta$\,Sct star HD\,184471 by \citet{Kudryavtsev2006}, who used the Main Stellar  
Spectrograph installed on the 6-m  telescope at the Special Astrophysical Observatory in Russia.
However, HD\,184471 is a member of a spectroscopic binary \citep{Mathys2017} and there has not been a study
 yet to confirm that the $\delta$\,Sct-like pulsations originate in the primary component.

It is conspicuous that both HD\,340577  and  HD\,184471 belong to the group of very slowly rotating Ap stars with 
rotation periods of 116.7\,d and 50.8\,d, respectively (\citealt{Mathys2022}, and references therein).
One more magnetic Ap star, HD\,191742, with $\bz$ between 0.6 and 1.8\,kG is reported by
\citet{Mathys2022} as a  $\delta$\,Sct candidate. 
Furthermore, \citet{Mathys2017} reported on two other Ap stars with very long periods, HD\,965 and HD\,55719,
which exhibit in their polarimetric spectra crossover that 
cannot be accounted for by rotation. These two stars also appear to be good candidates for the search of low- or
high-overtone $p$-mode pulsations. Since $\delta$\,Sct variables with magnetic fields are extremely rare among Ap stars,
it appears more appropriate to narrow the focus of the search for $\delta$\,Sct pulsations 
towards the sample of  Ap stars with long rotation periods.

HD\,340577 appears to be similar to typical magnetic Ap stars, apart from the rather low 
Pr and Nd abundances. Interestingly, long ago, \citet{Cowley1979} mentioned five stars 
characterized by strong lines of Fe-peak elements and REE lines
weaker than usually observed in Ap stars. Later, \citet{Koch2006} reported that also the Ap star 
HD\,133792 belongs to this group.
It is striking that low Pr and Nd abundances were in the past 
also reported for the two rapidly oscillating Ap stars $\beta$\,CrB and HD\,116114 with the lowest pulsation 
amplitudes and rather long pulsation periods, 16.2\,min and  21.2\,min, respectively
(see e.g.\ \citealt{Kurtz2007} and references therein).
\citet{Alentiev2012} compared the chemical composition of the roAp star HD\,177765 with the 
element abundances in the atmosphere of the roAp star $\beta$\,CrB (both with long pulsation periods) and 
reported that both stars exhibit a pronounced CeEu ionization anomaly but not the PrNd anomaly,
which is typical for roAp stars with shorter pulsation periods. 
Future studies of the chemical 
composition of HD\,340577 using high-resolution spectra with a better 
spectral coverage are necessary to confirm this trend also for this star.

The high-order $g$ modes commonly observed in $\gamma$\,Dor stars are expected to be heavily 
damped by magnetic fields
stronger than $1-4$\,kG, with the damping being stronger for higher radial orders
\citep{Murphy2020}.
No magnetic field detection has been reported in the past for any star with $\gamma$\,Dor pulsations. 
Also our measurements of the $\gamma$\,Dor star
HR\,8799 confirm the absence of a magnetic field with an upper limit of $\sim$9\,G. 
According to \citet{Murphy2020}, the strong damping in $\gamma$\,Dor can perhaps be attributed to the
dominant horizontal motions of their $g$ modes, which would strongly disturb magnetic field lines.

\section*{Acknowledgements}
We would like to thank the anonymous referee for their suggestions.
This work is based on observations carried out with the PEPSI spectropolarimeter.
PEPSI was made possible by funding through the State
of Brandenburg (MWFK) and the German Federal Ministry of Education and
Research  (BMBF)  through  their  Verbundforschung  grants  05AL2BA1/3  and
05A08BAC. LBT Corporation partners are
the University of Arizona on behalf of the Arizona university system;
Istituto Nazionale di  Astrofisica,  Italy;
 LBT  Beteiligungsgesellschaft,  Germany,  representing  the
Max-Planck Society, the Leibniz-Institute for Astrophysics Potsdam (AIP), and
Heidelberg  University; the  Ohio  State  University; and  the  Research  Corporation,
on behalf of the University of Notre Dame, the University of Minnesota and
the University of Virginia. 

\section*{Data Availability}

The PEPSI data can be obtained from the authors upon reasonable request.

\bsp	
\label{lastpage}

\begin{thebibliography}{99}

\bibitem[\protect\citeauthoryear{Alentiev et al.}{2012}]{Alentiev2012}
Alentiev D., Kochukhov O., Ryabchikova T., Cunha M., Tsymbal V., Weiss W., 2012, MNRAS, 421, L82

\bibitem[\protect\citeauthoryear{Babcock}{1958}]{Babcock1958}
Babcock H.~W.,
1958, ApJS, 3, 141

\bibitem[\protect\citeauthoryear{Balona, Krisciunas, \& Cousins}{1994}]{Balona1994}
Balona L.~A., Krisciunas K., Cousins A.~W.~J.,
1994, MNRAS, 270, 905

\bibitem[\protect\citeauthoryear{Contro et al.}{2016}]{Contro2016}
Contro B., Horner J., Wittenmyer R.~A., Marshall J.~P., Hinse T.~C.,
2016, MNRAS, 463, 191

\bibitem[\protect\citeauthoryear{Donati, Semel \& Rees}{1992}]{Donati1992}
Donati J.-F., Semel M., Rees D.~E.,
1992, \aap, 265, 669

\bibitem[\protect\citeauthoryear{Cowley \& Henry}{1979}]{Cowley1979}
Cowley C.~R., Henry R., 1979, ApJ, 233, 633

\bibitem[\protect\citeauthoryear{Donati et al.}{1997}]{Donati1997} 
Donati J.-F., Semel M., Carter B.~D., Rees D.~E., Collier Cameron A.,
1997, MNRAS, 291, 658

\bibitem[\protect\citeauthoryear{Faramaz et al.}{2021}]{Faramaz2021}
Faramaz V., et al.,
2021, AJ, 161, 271

\bibitem[\protect\citeauthoryear{Gaia Collaboration}{2016}]{Gaia2016}
Gaia Collaboration, et al.,
2016, \aap, 595, A1

\bibitem[\protect\citeauthoryear{Gaia Collaboration et al.}{2021}]{Gaia2021}
Gaia Collaboration, et al.,
2021, A\&A, 649, A1

\bibitem[\protect\citeauthoryear{Gray \& Corbally}{2002}]{Gray2002}
Gray R.~O., Corbally C.~J., 2002, AJ, 124, 989

\bibitem[\protect\citeauthoryear{Grigahc{\`e}ne et al.}{2005}]{Grigahcene2005}
Grigahc{\`e}ne A., Dupret M.-A., Gabriel M., Garrido R., Scuflaire R.,
2005, A\&A, 434, 1055

\bibitem[\protect\citeauthoryear{Hubrig et al.}{2018}]{Hubrig2018}
Hubrig S., J{\"a}rvinen S.~P., Madej J., Bychkov V.~D., Ilyin I., Sch{\"o}ller M., Bychkova L.~V.,
2018, MNRAS, 477, 3791

\bibitem[\protect\citeauthoryear{Hubrig et al.}{2021}]{Hubrig2021}
Hubrig S., J{\"a}rvinen S.~P., Ilyin I., Strassmeier K.~G., Sch{\"o}ller M.,
2021, MNRAS, 508, L17

\bibitem[\protect\citeauthoryear{Hubrig \& Sch{\"o}ller}{2021}]{HubrigSchoeller2021}
Hubrig S., Sch{\"o}ller M.,
2021, ``Magnetic Fields in O, B, and A Stars'', ISBN: 978-0-7503-2390-1. IOP ebooks. Bristol, UK: IOP Publishing

\bibitem[\protect\citeauthoryear{J{\"a}rvinen et al.}{2018}]{Jarvinen2018}
J{\"a}rvinen S.~P., Hubrig S., Scholz R.-D., Niemczura E., Ilyin I., Sch{\"o}ller M.,
2018, MNRAS, 481, 5163

\bibitem[\protect\citeauthoryear{Kennedy \& Wyatt}{2014}]{KennedyWyatt2014}
Kennedy G.~M., Wyatt M.~C.,
2014, MNRAS, 444, 3164

\bibitem[\protect\citeauthoryear{Kochukhov et al.}{2006}]{Koch2006}
Kochukhov O., Tsymbal V., Ryabchikova T., Makaganyk V., Bagnulo S., 2006, A\&A, 460, 831

\bibitem[\protect\citeauthoryear{Koen et al.}{2001}]{Koen2001}
Koen C., et al.,
2001, MNRAS, 326, 387

\bibitem[\protect\citeauthoryear{Kudryavtsev et al.}{2006}]{Kudryavtsev2006}
Kudryavtsev D.~O., Romanyuk I.~I., Elkin V.~G., Paunzen E.,
2006, MNRAS, 372, 1804

\bibitem[\protect\citeauthoryear{Kupka, Dubernet \& the VAMDC Collaboration}{2011}]{Kupka2011}
Kupka F., Dubernet M.-L., VAMDC Collaboration,
2011, Baltic Astronomy, 20, 503

\bibitem[\protect\citeauthoryear{Kurtz, Elkin, \& Mathys}{2007}]{Kurtz2007}
Kurtz D.~W., Elkin V.~G., Mathys G.,
2007, MNRAS, 380, 741

\bibitem[\protect\citeauthoryear{Kurtz et al.}{2008}]{Kurtz2008}
Kurtz D.~W., Hubrig S., Gonz{\'a}lez J.~F., van Wyk F., Martinez P.,
2008, MNRAS, 386, 1750

\bibitem[\protect\citeauthoryear{Li et al.}{2020}]{Li2020}
Li G., Van Reeth T., Bedding T.~R., Murphy S.~J., Antoci V., Ouazzani R.-M., Barbara N.~H.,
2020, MNRAS, 491, 3586

\bibitem[\protect\citeauthoryear{Renson \& Manfroid}{2009}]{Renson2009}
Renson P., Manfroid J., 2009, A\&A, 498, 961

\bibitem[\protect\citeauthoryear{Marois et al.}{2008}]{Marois2008}
Marois C., et al., 2008, Sci., 322, 1348

\bibitem[\protect\citeauthoryear{Mathys \& Lanz}{1992}]{MathysLanz1992}
Mathys G., Lanz T.,
1992, A\&A, 256, 169

\bibitem[\protect\citeauthoryear{Mathys, Kurtz, \& Holdsworth}{2022}]{Mathys2022}
Mathys G., Kurtz D.~W., Holdsworth D.~L.,
2022, A\&A, 660, A70

\bibitem[\protect\citeauthoryear{Mathys}{2017}]{Mathys2017}
Mathys G.,
2017, A\&A, 601, A14

\bibitem[\protect\citeauthoryear{Murphy et al.}{2020}]{Murphy2020}
Murphy S.~J., Saio H., Takada-Hidai M., Kurtz D.~W., Shibahashi H., Takata M., Hey D.~R.,
2020, MNRAS, 498, 4272

\bibitem[\protect\citeauthoryear{Neiner, Wade, \& Sikora}{2017}]{Neiner2017}
Neiner C., Wade G.~A., Sikora J.,
2017, MNRAS, 468, L46

\bibitem[\protect\citeauthoryear{Reidemeister et al.}{2009}]{Reidemeister2009}
Reidemeister M., Krivov A.~V., Schmidt T.~O.~B., Fiedler S., M{\"u}ller S., L{\"o}hne T., Neuh{\"a}user R.,
2009, A\&A, 503, 247

\bibitem[\protect\citeauthoryear{Robrade \& Schmitt}{2010}]{RobradeSchmitt2010}
Robrade J., Schmitt J.~H.~M.~M.,
2010, A\&A, 516, A38

\bibitem[\protect\citeauthoryear{Saio}{2005}]{Saio2005}
Saio H.,
2005, MNRAS, 360, 1022

\bibitem[\protect\citeauthoryear{S{\'o}dor et al.}{2014}]{Sodor2014}
S{\'o}dor {\'A}., et al.,
2014, A\&A, 568, A106

\bibitem[\protect\citeauthoryear{Strassmeier et al.}{2015}]{Strassmeier2015}
Strassmeier K.~G., et al.,
2015,  Astron.\ Nachr., 336, 324

\bibitem[\protect\citeauthoryear{Strassmeier, Carroll, \& Ilyin}{2023}]{Strassmeier2023}
Strassmeier K.~G., Carroll T.~A., Ilyin I.~V.,
2023, arXiv:2305.07470

\bibitem[\protect\citeauthoryear{Su et al.}{2009}]{Su2009}
Su K.~Y.~L., et al.,
2009, ApJ, 705, 314

\bibitem[\protect\citeauthoryear{Su et al.}{2013}]{Su2013}
Su K.~Y.~L.,  et al.,
2013, ApJ, 763, 118

\bibitem[\protect\citeauthoryear{Wright et al.}{2011}]{Wright2011}
Wright D.~J., et al.,
2011, ApJL, 728, L20

\bibitem[\protect\citeauthoryear{Zwintz et al.}{2020}]{Zwintz2020}
Zwintz K., et al.,
2020, A\&A, 643, A110

\end{thebibliography}
\end{document}